\documentclass[12pt]{iopart}
\usepackage{iopams}
\usepackage{graphicx}

\begin{document}

\title[Thermal expansion and magnetostriction of YNi$_2$B$_2$C]{%
Anisotropic thermal expansion and magnetostriction of YNi$_2$B$_2$C single crystals}

\author{S.~L. Bud'ko\P, G.~M. Schmiedeshoff\dag, G. Lapertot\P\ddag, and P.~C. Canfield\P}

\address{\P\ Ames Laboratory US DOE and Department of Physics and Astronomy,
Iowa State University, Ames, IA 50011, USA}
\address{\dag\ Department of Physics,
Occidental College, Los Angeles, CA 90041, USA}
\address{\ddag\ Commissariat \'a l'Energie Atomique, DRFMC-SPSMS-IMAPEC, 38054 Grenoble, France}

\begin{abstract}
We present results of anisotropic thermal expansion and low temperature magnetostriction measurements on
YNi$_2$B$_2$C single crystals grown by high temperature flux and floating zone techniques. Quantum oscillations of
magnetostriction were observed at low temperatures for $H \| c$ starting at fields significantly below $H_{c2}$
($H < 0.7 H_{c2}$). Large irreversible, longitudinal magnetostriction was seen in both, in-plane and along the
$c$-axis, directions of the applied magnetic field in the intermediate superconducting state. Anisotropic uniaxial
pressure dependencies of $T_c$ were evaluated using results of zero field, thermal expansion measurements.
\end{abstract}


\submitto{\JPCM}

\maketitle

\section{Introduction}
The members of the $R$Ni$_2$B$_2$C ($R$ = Gd-Lu, Y) series of compounds serve as model systems for studies of a
number of phenomena: co-existence of local moment magnetism and superconductivity, non-locality and flux line
lattice transitions, heavy fermion physics and complex metamagnetism \cite{can98a,mul01a,mul02a,bud06a}. Although
many sophisticated experiments were performed on $R$Ni$_2$B$_2$C borocarbides, data on thermal expansion (TE) and
magnetostriction (MS) for several members of the series exist only for limited temperature ranges, or only on
polycrystalline samples, or are not available at all. For one of the less complex members of the series,
non-magnetic, superconducting YNi$_2$B$_2$C, for example, only a rather limited set of results has been
disseminated \cite{ful98a,bel98a,jae02a}. In this publication we report anisotropic ($c$-axis and $ab$-plane) TE
and longitudinal, low temperature, MS measurements on  YNi$_2$B$_2$C single crystals grown by two different
techniques and well-characterized by other methods. The objective for this work is manifold: to acquire data that
can serve as a baseline in studies of TE in more complex, magnetic or strongly correlated, borocarbides; to probe
the irreversible properties in intermediate superconducting state; and finally, to evaluate the uniaxial pressure
derivatives of YNi$_2$B$_2$C from the anomalies in TE at $T_c$ (using the the thermodynamic Ehrenfest relation),
an approach proven viable {\it e.g.} in HTSC \cite{mei90a,gug94a}, and compare the results with the implications
of the analysis in Ref. \cite{san00a}.

\section{Experimental methods}
Two YNi$_2$B$_2$C single crystals were used in this work. One of them (sample A throughout the rest of the text)
was grown by a Ni$_2$B high temperature flux method (see Ref. \cite{can98a,mxu94a,can01a} for more details). This
method yields plate-like crystals with the $c$-axis perpendicular to the plates. The crystal was shaped into a
nearly rectangular bar, with two pairs of parallel surfaces, so that TE and MS were measured along [100] ($L =
2.37$ mm) and [001] ($L = 0.92$ mm) directions. Another crystal (sample B) was grown using vertical zone melting
method with a commercial 4-mirrors image furnace (model FZ-T-4000-H-VI-VPM-PC, Crystal Systems Corp., Japan). We
used drop cast polycrystalline rods ($\sim 6$ mm diameter, $\sim 60$ mm long) as feeding rod and crystal support.
High purity argon at atmospheric pressure was used as a protective atmosphere inside the quartz working tube.
Feeding and bottom shafts were rotated in opposite directions to insure an effective stirring of the molten zone,
typically at +10 and -35 rpm. During the growth, the growth speed was decreased from 10 to 2 mm/h to favor the
formation of large, single grains. The phase purity of the single crystals was checked by Debye-Sherrer powder
diffraction pattern and compared to original feeding rod spectrum. No detectable extra reflections due to impurity
phases or changes of cell parameters were observed. An as-grown bar was oriented with a Laue camera and cut into a
close to prismoidal shape with two pairs of parallel surfaces, so that TE and MS were measured along [110] ($L =
3.03$ mm) and [001] ($L = 1.82$ mm) directions. After shaping, both samples were annealed in dynamic vacuum
($10^{-5}-10^{-6}$ Torr) at 950$^\circ$ C \cite{mia02a}. It should be mentioned that for reasons unrelated to the
objectives of this work, the sample A was grown with isotopically pure $^{10}$B and the sample B with $^{11}$B.

Magnetization measurements were performed using a commercial MPMS-5 (Quantum Design, Inc.) SQUID magnetometer; ac
resistance ($f = 16$ Hz, $I = 3$ mA) and heat capacity were measured with ACT and heat capacity options of a
PPMS-9 instrument (Quantum Design, Inc.).

Thermal expansion and magnetostriction were measured using a capacitive dilatometer constructed of OFHC copper; a
detailed description of the dilatometer will appear elsewhere \cite{sch06a}.  The capacitance was measured with an
Andeen-Hagerling 2500A capacitance bridge (the bridge resolution of 10$^{-7}$ pF corresponds to a sample dilation
measurement limit of about 0.3 pm when the dilatometer is operating near 20 pF).  The dilatometer was mounted in a
PPMS-14 instrument (Quantum Design Inc.) and was operated over a temperature range of 1.8 to 305 K and in magnetic
fields up to 140 kOe (either in vacuum or with a small amount of helium exchange gas to minimize thermal
gradients). The temperature was measured with a calibrated Cernox-1030 thermometer (Lakeshore Cryotronics) mounted
on the dilatometer.  The field dependence of the dilatometer was less than $\pm 3 $ \AA\enspace over $\pm 140$ kOe
and will be ignored.  The temperature dependence of the dilatometer (or ``cell effect'') is removed using
published values of the thermal expansion of copper \cite{kro77a}. Data were acquired with the temperature
increasing at a rate of about 0.4 K/min or with the field changing at a rate of about 6 kOe/min.  The absolute
accuracy of the dilatometer was checked with measurements on a 4 mm sample of pure aluminum that were compared to
published values \cite{kro77a}. The average deviation between our aluminum TE measurements and those in the
literature over our full temperature range is $7.0\times{10}^{-8}$ K$^{-1}$.  The average fractional deviation of
our aluminum measurements from those in the literature is about 1\% above 40 K but becomes larger at low
temperatures where the thermal expansion of aluminum is sensitive to both sample purity and preparation
\cite{kro77a}.  The maximum deviation of $2.8\times{10}^{-7}$ K$^{-1}$ occurred at 300 K and corresponds to a
fractional deviation of 1.2\% from published values. We estimate an experimental uncertainty of ${10}^{-8}$
K$^{-1}$ in our thermal expansion measurements from the mean deviation to a fit of our aluminum data near 20 K,
this corresponds to a dilation uncertainty less than 0.5 \AA/K.

\section{Results and Discussion}
\subsection{Thermal expansion}
\begin{figure}[tbp]
\begin{center}
\includegraphics[angle=0,width=120mm]{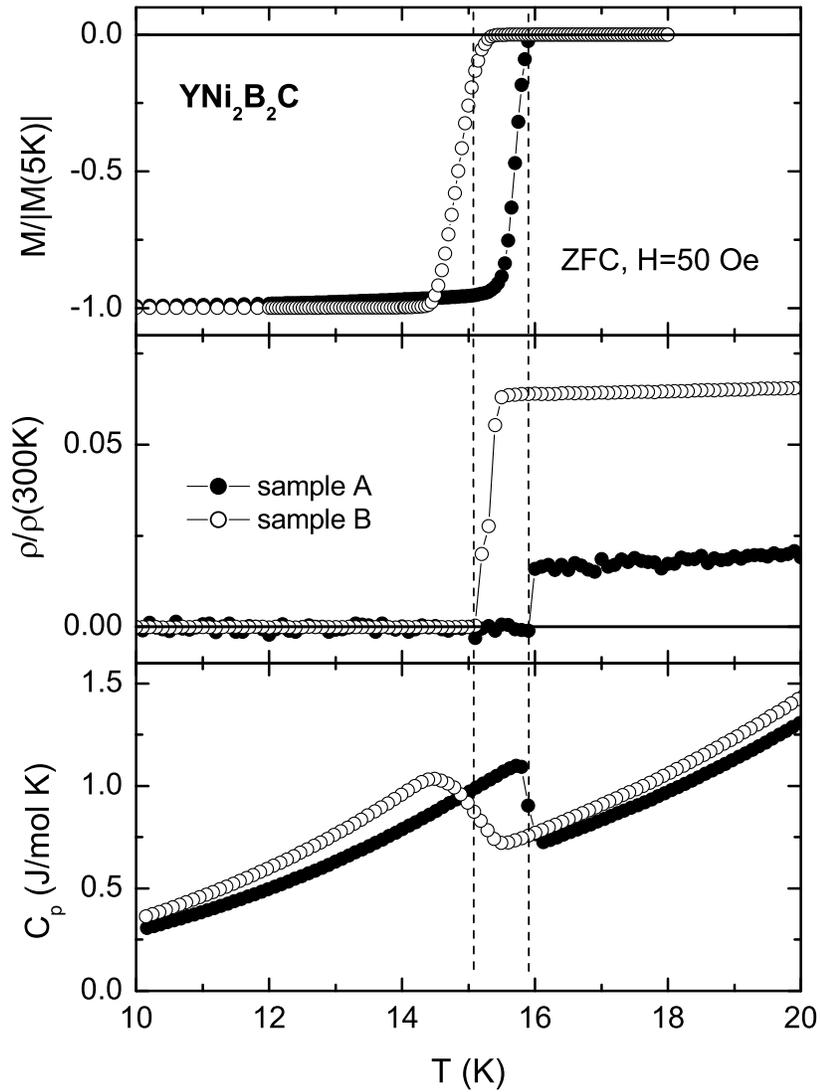}
\end{center}
\caption{Temperature-dependent magnetization, resistivity and heat capacity for samples A and B near
superconducting transition. Vertical lines mark transition as defined by $\rho = 0$.} \label{F1}
\end{figure}

Superconducting transitions, as measured by DC magnetization, zero field resistivity and heat capacity, are shown
in Fig. \ref{F1}. In all three measurements the transitions are sharp, being sharper for sample A. The $T_c$
values as defined by $\rho = 0$, onset of diamagnetism in $M(T)$ measurements and mid-point of the heat capacity
rise (or by balancing of the normal state and superconducting state entropies) are $\sim 15.9$ K and $\sim 15.1$ K
for samples A and B respectively. This difference in $T_c$ between samples A and B is partially due to the boron
isotope effect \cite{law94a,che99a} and partially due to different scattering in two samples as can be see from
the difference in the residual resistivity ratio, $RRR = \rho(300$K$)/\rho(17$K$) \approx$ 63 and 16 for A and B
respectively \cite{mia02a} . It should be noted that sample B had a small, rather broad, bump in resistivity near
30 K, feature that was not observed in the sample A or other flux-grown YNi$_2$B$_2$C samples.

\begin{figure}[tbp]
\begin{center}
\includegraphics[angle=0,width=120mm]{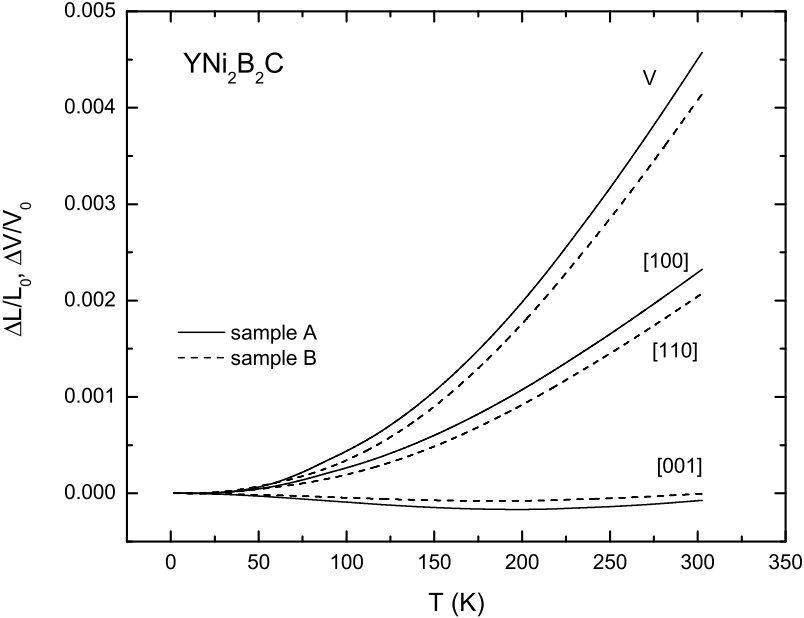}
\end{center}
\caption{Relative temperature-dependent linear and volume dilations of two YNi$_2$B$_2$C crystals.} \label{F2}
\end{figure}

Temperature-dependent linear and volume (defined here as $\Delta V/V_0 = \Delta L_c/L_{c0} + 2 \Delta
L_{ab}/L_{ab0}$) dilations (relative to the values at 1.9 K) for two samples are shown in Fig. \ref{F2}. The two
crystals behave very similarly with the slight discrepancy most probably reflecting the accuracy of the
measurements of the room temperature dimensions and their changes and possible, slight imperfections in the shape
of the samples. In-plane linear and volume thermal dilations are positive in the temperature range studied. The
$c$-axis thermal dilation is negative at low temperatures, it changes its sign at about 200 K.

\begin{figure}[tbp]
\begin{center}
\includegraphics[angle=0,width=120mm]{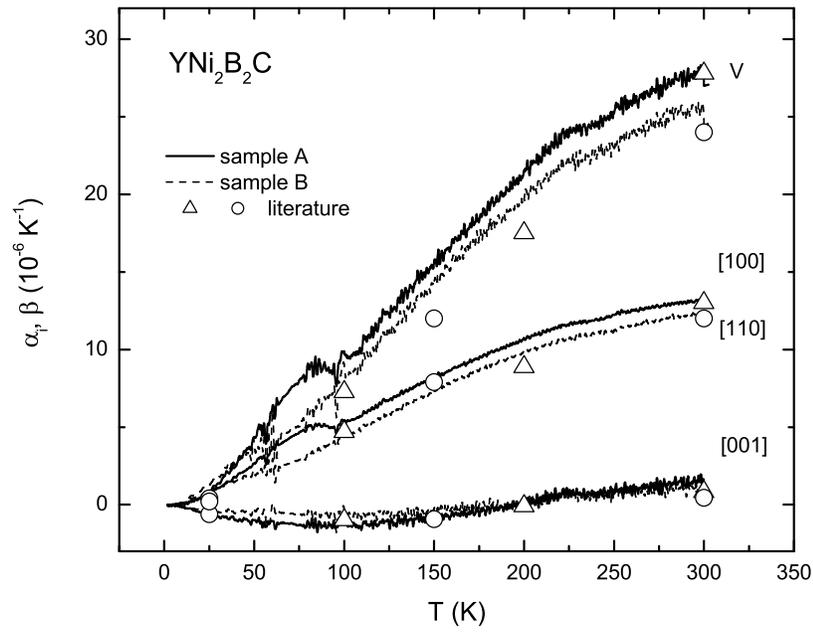}
\end{center}
\caption{Temperature-dependent linear and volume thermal expansion coefficients for two YNi$_2$B$_2$C crystals
plotted together with literature data; symbols: $\bigcirc$ - Ref. \cite{bel98a}, $\bigtriangleup$ - Ref.
\cite{jae02a} (see text for discussion of the literature values).} \label{F3}
\end{figure}

Temperature-dependent linear and volume thermal expansion coefficients measured on samples A and B are shown in
Fig. \ref{F3}. $\alpha_c(T)$ data for both samples are very similar, as well as the $\alpha_{ab}(T)$ data above
$\sim 100$ K. Further studies will be required to understand whether the broad maximum in $\alpha_{[100]}(T)$ of
the sample A at about 80 K is a real feature albeit it was reproducible in two consecutive measurements:, one in
high vacuum mode of the PPMS, another with low pressure He exchange gas in the sample chamber. Our data are
compared with the published data \cite{bel98a,jae02a} obtained by powder and single crystal X-ray diffraction at
different temperatures. To be consistent with the Fig. 3 and results for $\beta$ in the Table 3 of Ref.
\cite{jae02a}, the  $\alpha_c$ data cited in ref. \cite{jae02a} for YNi$_2$B$_2$C should be divided by 10. Such
"corrected" $\alpha_c$ data together with the $\alpha_a$ and $\beta$ data from the Table 3, Ref. \cite{jae02a} are
included in Fig. \ref{F3}. Our data are consistent with the published values \cite{bel98a,jae02a}.

\begin{figure}[tbp]
\begin{center}
\includegraphics[angle=0,width=120mm]{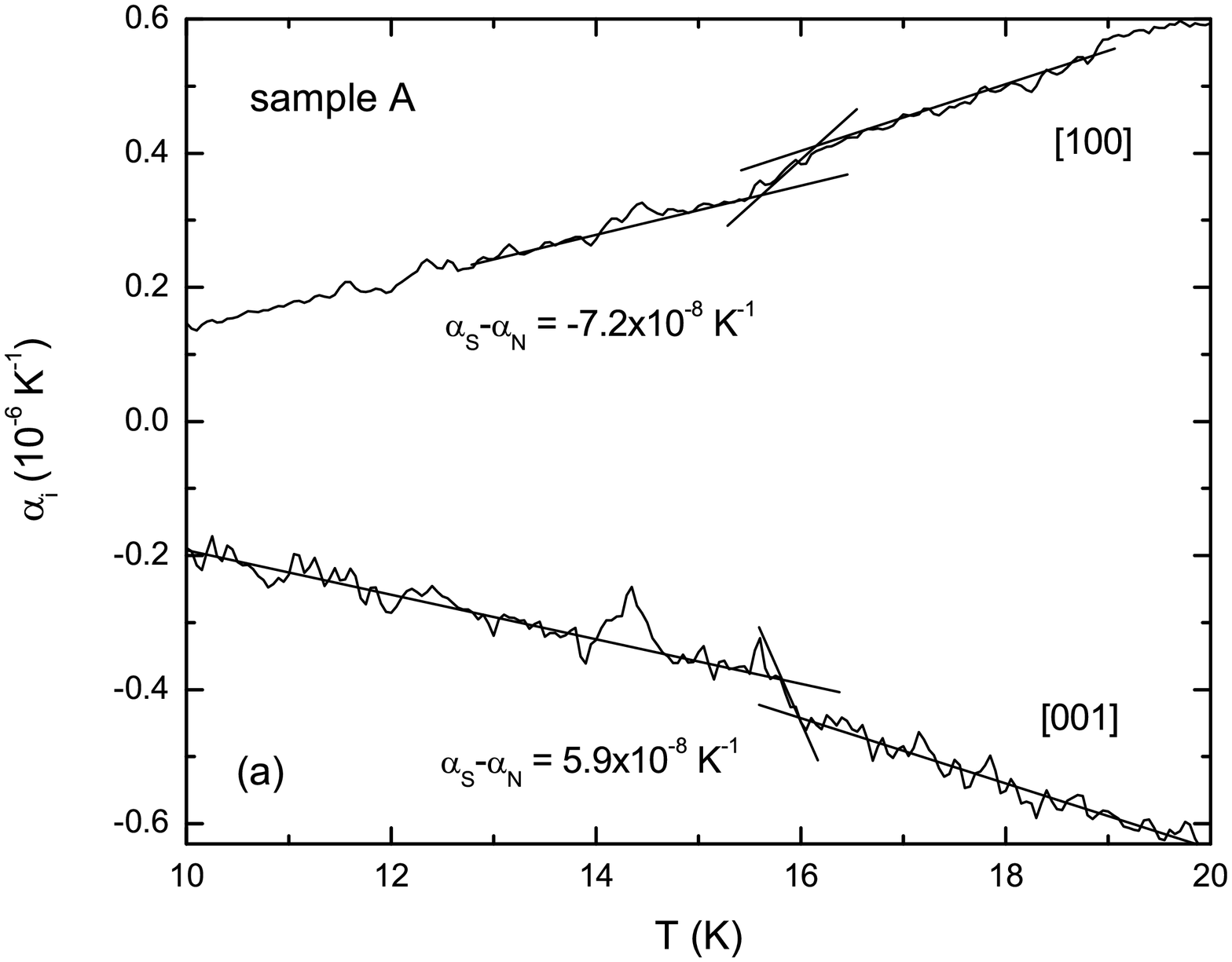}
\includegraphics[angle=0,width=120mm]{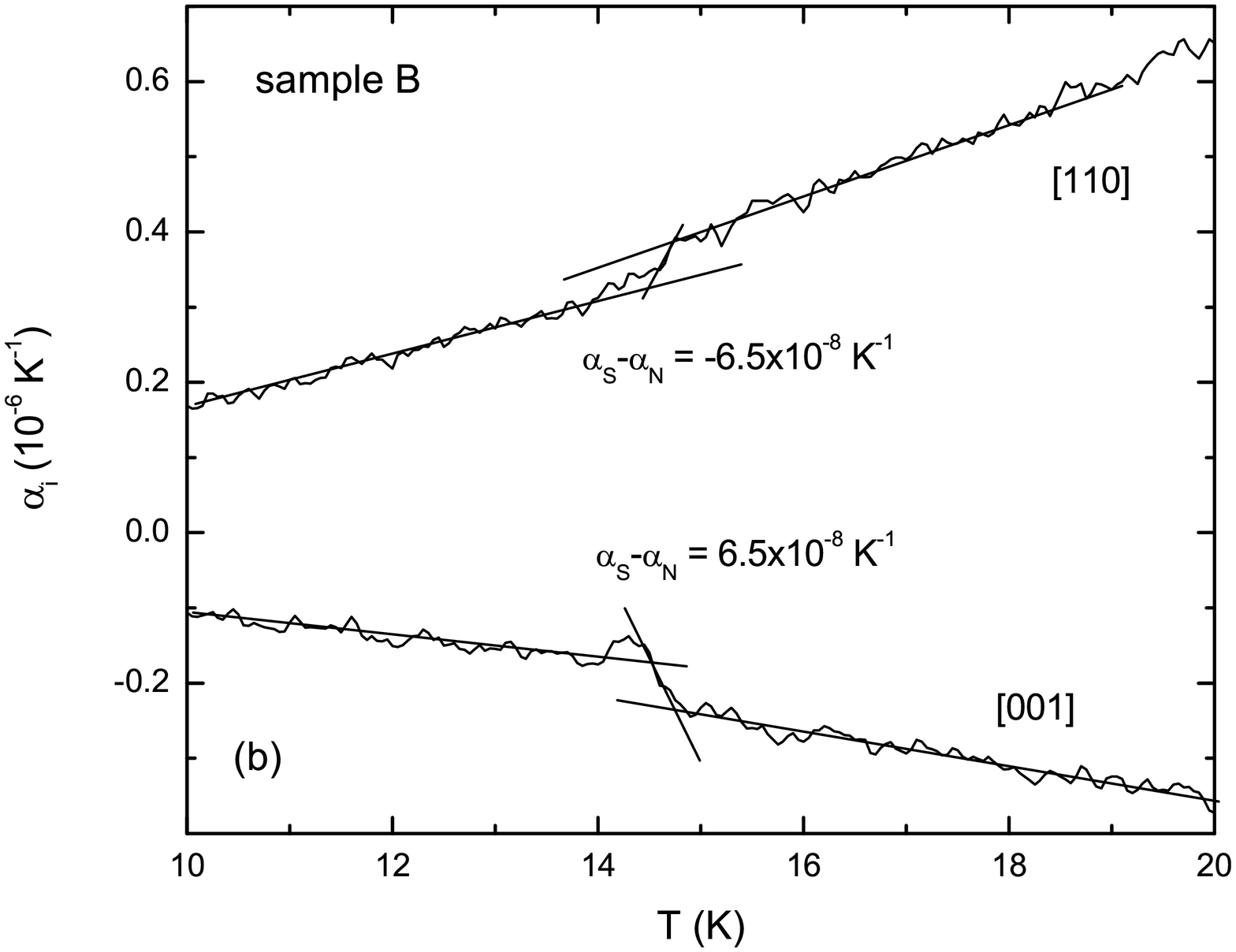}
\end{center}
\caption{Temperature-dependent linear thermal expansion coefficients for two YNi$_2$B$_2$C crystals in the
vicinity of the superconducting transition.} \label{F4}
\end{figure}

The temperature dependence of the linear thermal expansion coefficients close to $T_c$ is presented in Fig.
\ref{F4}. Jumps in $\alpha_i(T)$ are seen unambiguously, with the opposite sign of the jumps for $\alpha_{ab}$ and
$\alpha_c$ and very similar, within the errors of the measurements, values for both samples. Using results from
the heat capacity (Fig. \ref{F1}) and thermal expansion coefficients (Fig. \ref{F4}) measurements, the uniaxial
pressure derivatives of the superconducting transition temperature, $T_c$, can be calculated using the
thermodynamic Ehrenfest relation:

\begin{displaymath}
\frac {dT_c}{dp_i} = V_{mol} \cdot \Delta \alpha_i(T_c) \cdot \left[ \frac{\Delta C_p(T_c)}{T_c} \right] ^{-1}
\end{displaymath}

\noindent where $V_{mol}$ is a molar volume of the material (for YNi$_2$B$_2$C, $V_{mol} = 39.5$ cm$^3$) and
$\Delta \alpha_i(T_c)$ and $\Delta C_p(T_c)$ are the jumps in the $i$-th thermal expansion coefficient and
specific heat at $T_c$. Estimates of the uniaxial pressure derivatives for YNi$_2$B$_2$C based on the
aforementioned measurements are summarized in Table \ref{tab}. Within the 10-20\% error bars in $\Delta \alpha_i$
(see Fig. \ref{F4}) there is no significant difference in estimated uniaxial pressure derivatives for two samples.
In the table $dT_c/dP* = 2 \cdot dT_c/dp_{ab} + dT_c/dp_c$. The $dT_c/dP*$ defined in such way lacks the
contribution from the off-diagonal $dT_c/dp_{ij}$ terms that play a role in the experimentally measured $dT_c/dP$
under hydrostatic pressure. Since the off-diagonal terms are usually significantly smaller then the diagonal ones,
we can still compare the last column of the Table \ref{tab} with the experimentally measured hydrostatic pressure
derivatives. The experimental data on $dT_c/dP$ of YNi$_2$B$_2$C were presented in a number of publications, some
of the reported values are listed below (in K/kbar): $-0.58 \cdot 10^{-2}$ \cite{sch94a}, $-0.9 \cdot 10^{-2}$
\cite{mur94a}, $0.32 \cdot 10^{-2}$ (in $P \to 0$ limit) \cite{all95a}, and $-0.9 \cdot 10^{-2}$ \cite{loo95a}.
All but one of these results are consistent with our estimates. We are unaware of any direct uniaxial pressure
measurements on YNi$_2$B$_2$C other than very short conference proceedings publication \cite{kob06a}, that
contains a statement that for this material $T_c$ is almost independent on (uniaxial) pressure, again in agreement
with our estimates of $dT_c/dp_i$ (Table \ref{tab}).

The uniaxial pressure derivatives obtained from the Ehrenfest relations in this work differ in value and, more
importantly, have the signs opposite to the ones inferred in Ref. \cite{san00a}. The procedure used there was
indirect, based on a number of assumptions (the most important, but not universally correct, being equivalence of
physical and chemical pressure), had to rely (due to the nature of the doping used) on refining of the structural
data of two-phase samples, and therefore is open for discussions and criticism. On the other hand, it is
noteworthy that additionally our current results (Table \ref{tab}) apparently contradict the implications of the
band-structure calculations \cite{mat94a} that superconductivity in RNi$_2$B$_2$C is controlled by the static
tetrahedral NiB$_4$ geometry. Within the model of Ref. \cite{mat94a} superconductivity is more favorable for
RNi$_2$B$_2$C compounds with the B-Ni-B tetrahedral angle $\phi$ closer to the ideal value of $\phi_{ideal} =
109.5^\circ$. For all RNi$_2$B$_2$C compounds $\phi < \phi_{ideal}$ (for YNi$_2$B$_2$C $\phi \approx 107.3^\circ$
\cite{bel98a}). Then a compression along the $c$-axis will cause a decrease in $\phi$ bringing it further away
from the $\phi_{ideal}$, and in-$ab$ plane compression will yield an increase in $\phi$ so that it will approach
$\phi_{ideal}$, consequently the signs of the uniaxial pressure derivatives of $T_c$ expected from \cite{mat94a}
are $dT_c/dp_c < 0$, $dT_c/dp_{ab} > 0$, opposite to that evaluated from the TE experiment (Table \ref{tab}).

\subsection{Magnetostriction}

\begin{figure}[tbp]
\begin{center}
\includegraphics[angle=0,width=120mm]{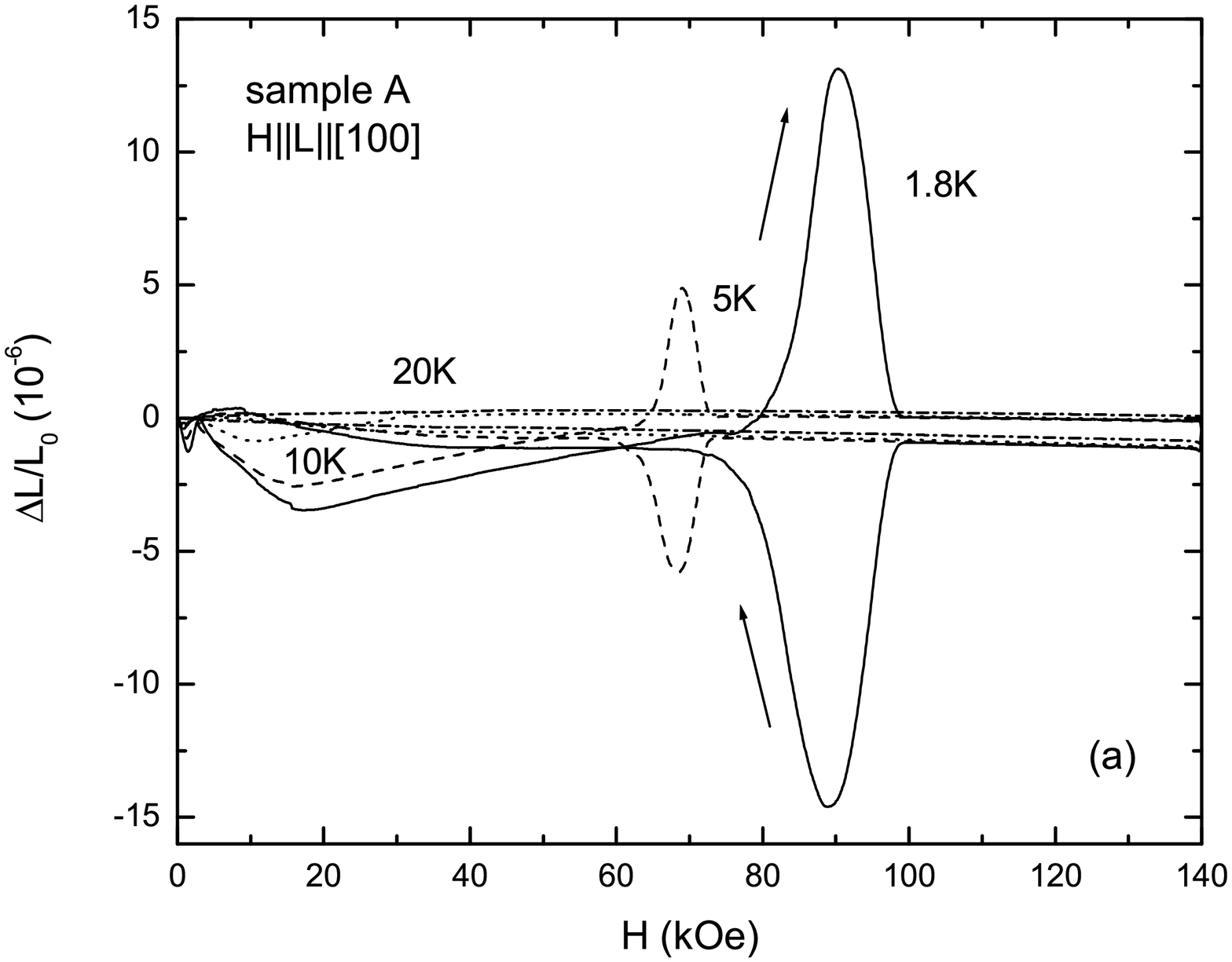}
\includegraphics[angle=0,width=120mm]{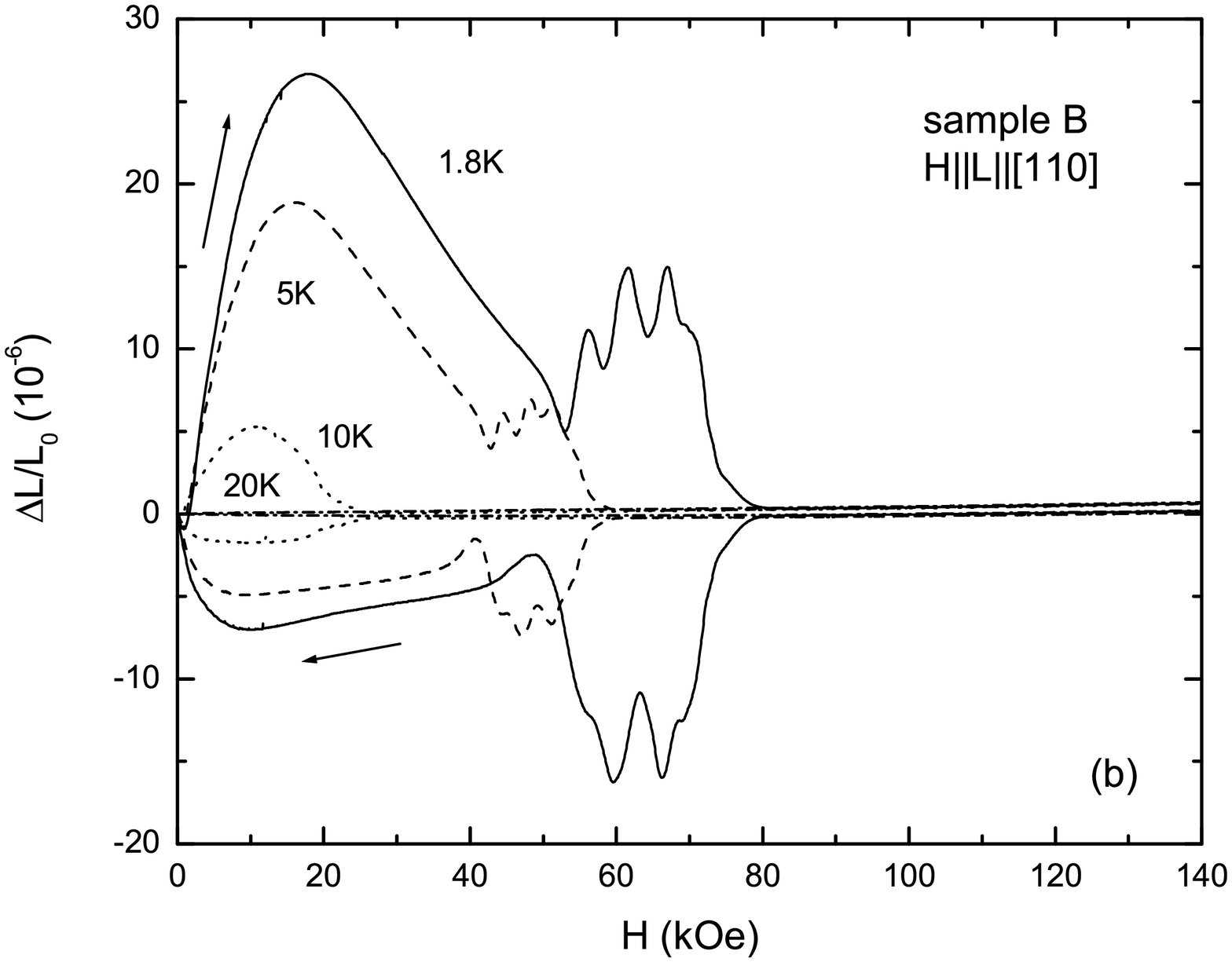}
\end{center}
\caption{Longitudinal magnetostriction for two YNi$_2$B$_2$C crystals with magnetic fields applied in the
$ab$-plane. $\Delta L$ is defined as $L(H) - L(H=0)$.} \label{F5}
\end{figure}

Magnetostriction loops taken at different temperatures, both in superconducting and normal state, for field
applied in the $ab$ plane, are shown in Fig. \ref{F5}. Magnetostriction in superconducting state depends on number
of parameters, including the extrinsic ones, like shape of the sample, pinning strength and its field dependence
\cite{iku94a,ere99a,joh00a,ger01a}. Moreover, irreversible flux pinning induced magnetostriction causes
(geometry-dependent) shape distortions \cite{joh00a,joh98a} that add additional requirements on mounting of the
sample in the experimental cell for detailed study of the critical superconducting state via magnetostriction.
Here we will just mention that in both samples the magnetostriction in the intermediate superconducting state is
high ($\Delta L_{max}/L_0 > 1 \cdot 10^{-7}$), significantly higher than that seen for the polycrystalline sample
\cite{ful98a}, in both samples the features associated with peak effect \cite{ger01a} or "dip" in magnetization
\cite{kog06a} are apparently seen, with some structure for the sample B. In the normal state magnetostriction is
rather small. In superconducting state, for magnetic fields appreciably lower than $H_{c2}$, the irreversible
magnetostriction is much higher for the sample B, consistent with its lower $RRR$ (Fig. \ref{F1}) and higher
pinning in this sample.

\begin{figure}[tbp]
\begin{center}
\includegraphics[angle=0,width=120mm]{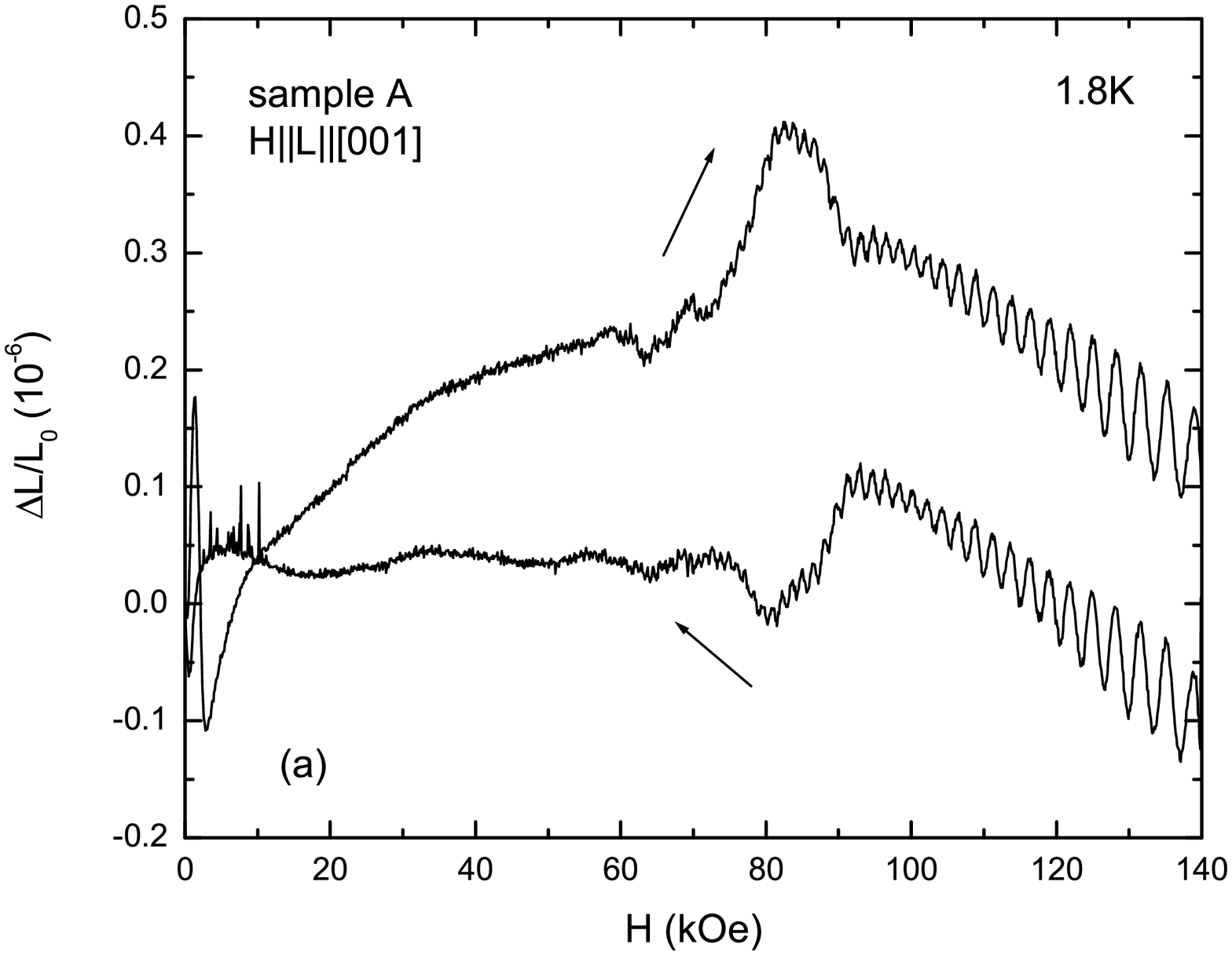}
\includegraphics[angle=0,width=120mm]{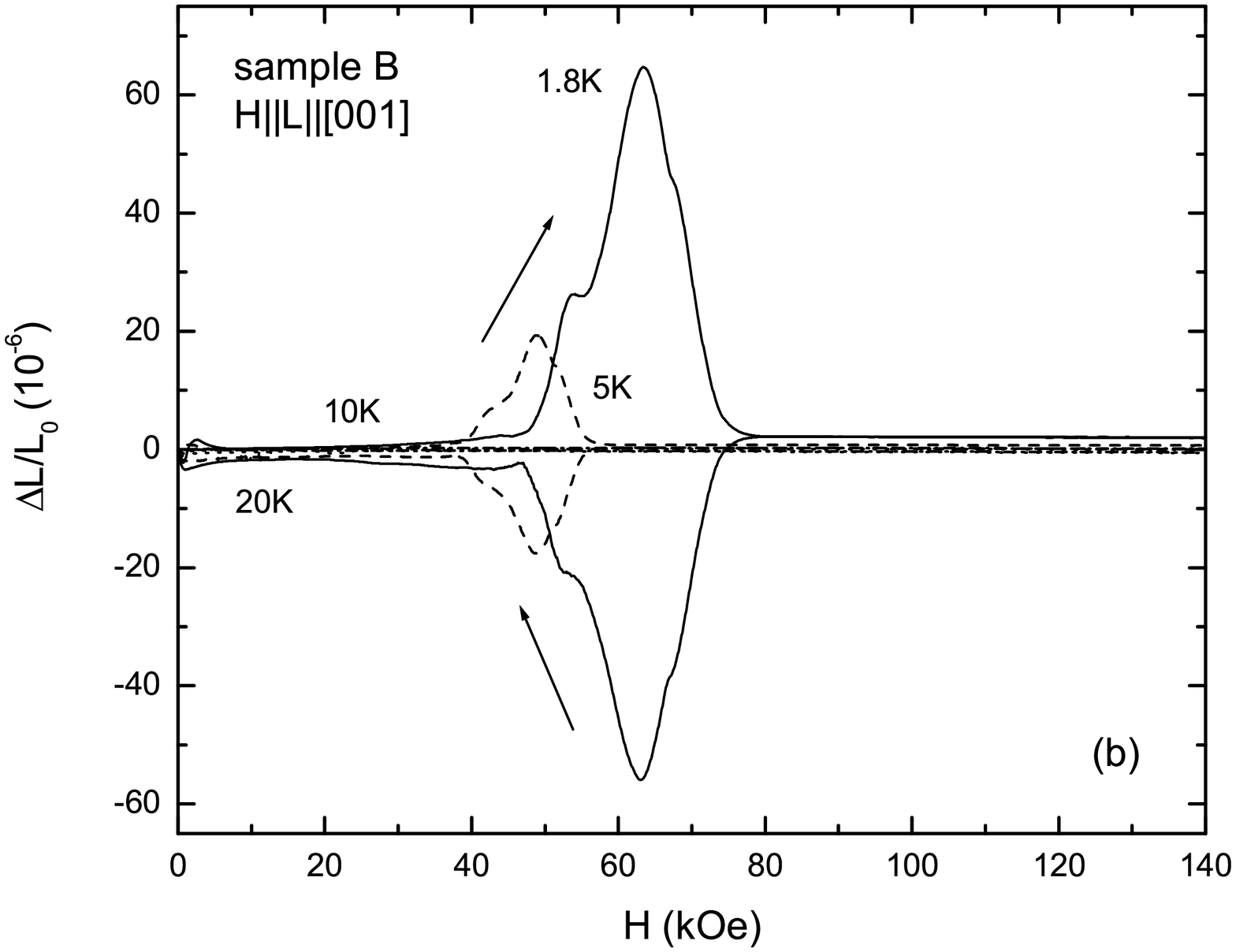}
\end{center}
\caption{Longitudinal magnetostriction for two YNi$_2$B$_2$C crystals with magnetic fields applied along the
$c$-axis. For sample A only data at $T = 1.8$ K are shown. $\Delta L$ is defined as $L(H) - L(H=0)$.} \label{F6}
\end{figure}

For field applied along $c$-direction the quantitative difference in the irreversible magnetostriction in
superconducting state between two samples is even more drastic that for $H \| ab$ (Fig. \ref{F6}). Very large,
($\Delta L_{max}/L_0 \approx 6 \cdot 10^{-7}$) effect is seen for sample B and is possibly associated with the
peak (and/or "dip") effect. Much smaller feature in the irreversible magnetostriction is seen below $H_{c2}$ in
the sample A.

\begin{figure}[tbp]
\begin{center}
\includegraphics[angle=0,width=120mm]{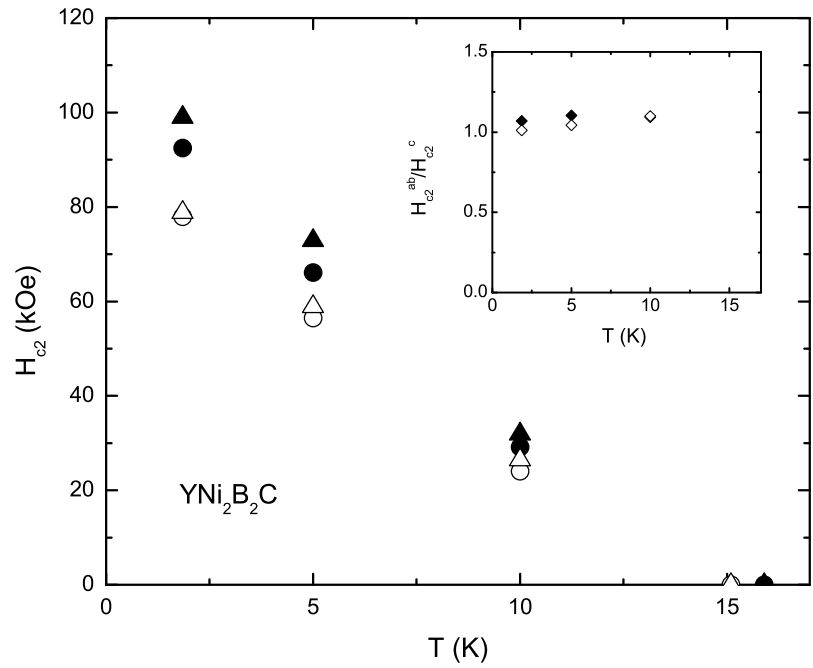}
\end{center}
\caption{Temperature dependent $H_{c2}$ for two YNi$_2$B$_2$C crystals as defined from magnetostriction
measurements. Triangles - $H \| ab$, circles - $H \| c$, filled symbols - sample A, open symbols - sample B.
Insert shows anisotropy of $H_{c2}$.} \label{Fad}
\end{figure}

The magnetostriction data allow for an estimate of the $H_{c2}$ (defined here as the high field onset of the
peak/dip effect feature). The anisotropic data for both samples are shown in Fig. \ref{Fad}. The values of
$H_{c2}$ and its anisotropy ($H_{c2}^{ab}/H_{c2}^c \sim 1.05-1.1$) are consistent with the previous results
obtained from magnetization and magnetoresistance measurements \cite{mxu94a,mar98a,bud01a,wim04a}, being on the
lower end of the reported values for the $H_{c2}$ anisotropy. The observation that $H_{c2}$ is lower for the
sample with lower $RRR$ is broadly consistent with the results \textit{e.g.} on Y(Ni$_{1-x}$Co$_x$)$_2$B$_2$C
\cite{che98a} and (Y$_{1-x}$Lu$_x$)Ni$_2$B$_2$C \cite{rat03a}, however it disagrees with the generally expected
increase of $H_{c2}$ with decrease of the mean free path (decrease of $RRR$) expected for conventional
superconductors (see Ref. \cite{hel66a} and refs therein). Careful study, by several techniques, of $H_{c2}$ and
its anisotropy in RNi$_2$B$_2$C and in particular in YNi$_2$B$_2$C as a function of its mean free path is a topic
of interest for a separate study.

\begin{figure}[tbp]
\begin{center}
\includegraphics[angle=0,width=100mm]{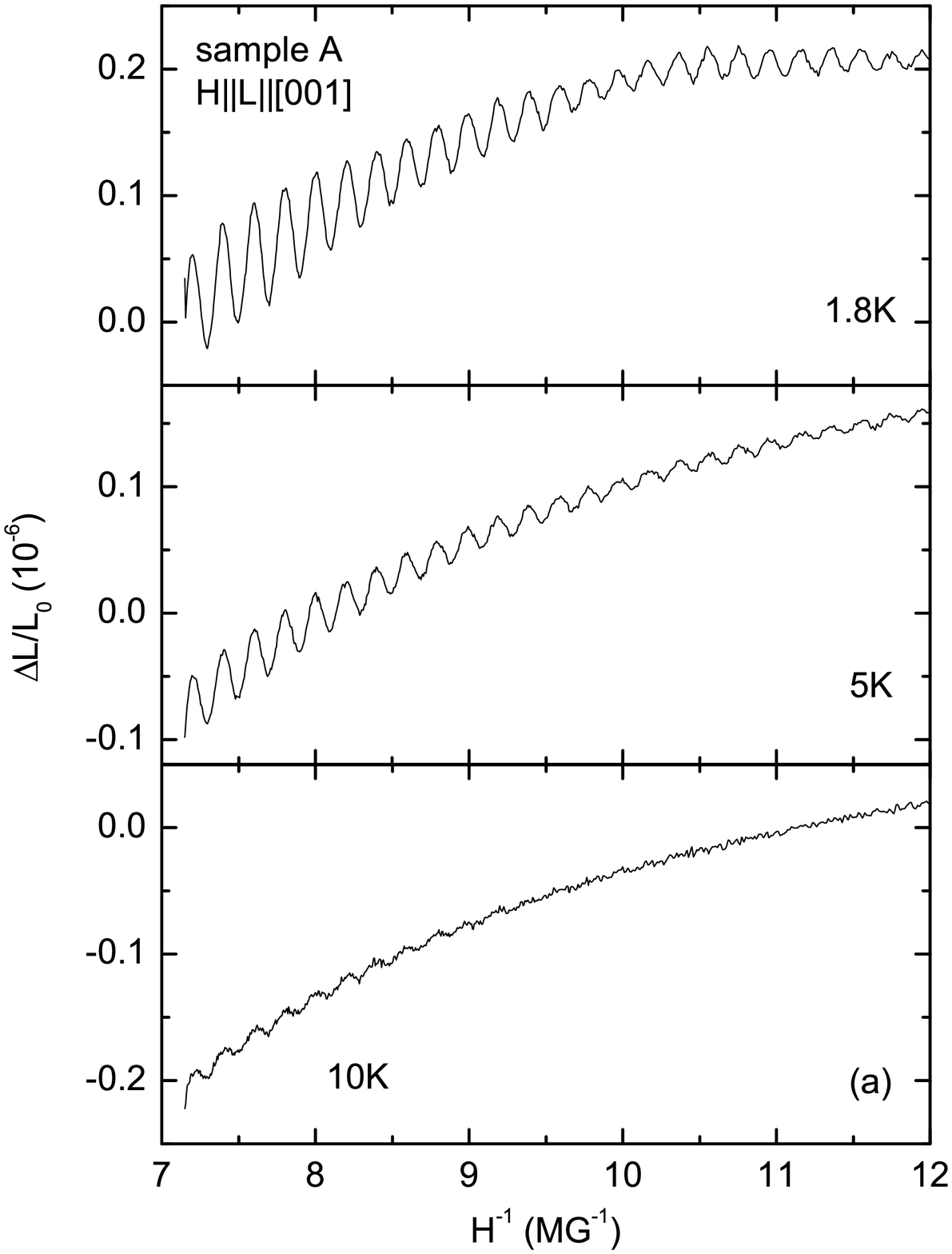}
\includegraphics[angle=0,width=100mm]{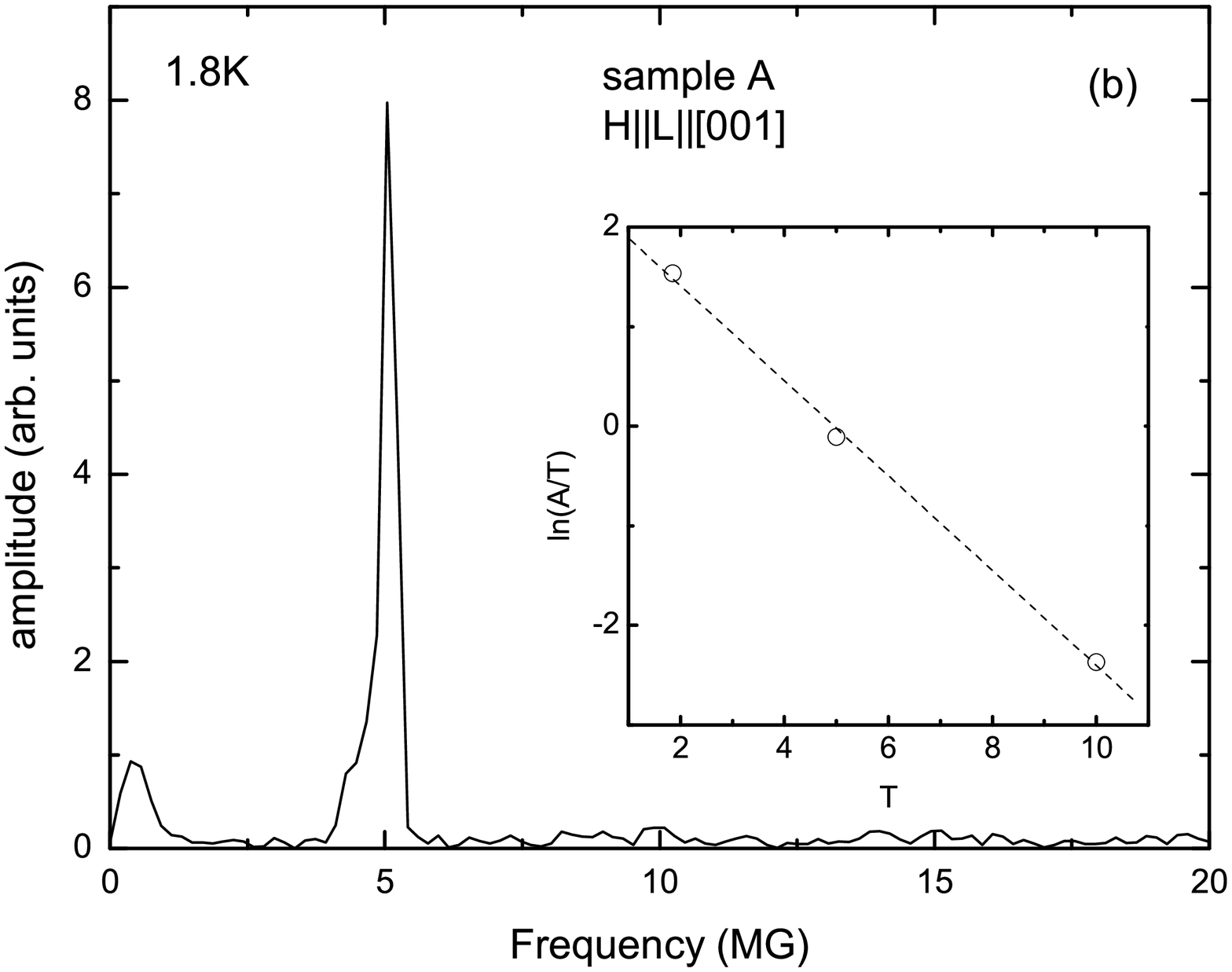}
\end{center}
\caption{(a) Quantum oscillations of the longitudinal magnetostriction plotted as a function of $1/H$ for sample
A. (b) Fourier transform of the $\Delta L/L_0$ vs. $1/H$ curve for $T = 1.8$ K. Inset: $ln(A/T)$ vs. $T$ plot for
amplitude $A$ at $H \sim 133$ kOe.} \label{F7}
\end{figure}

\begin{figure}[tbp]
\begin{center}
\includegraphics[angle=0,width=100mm]{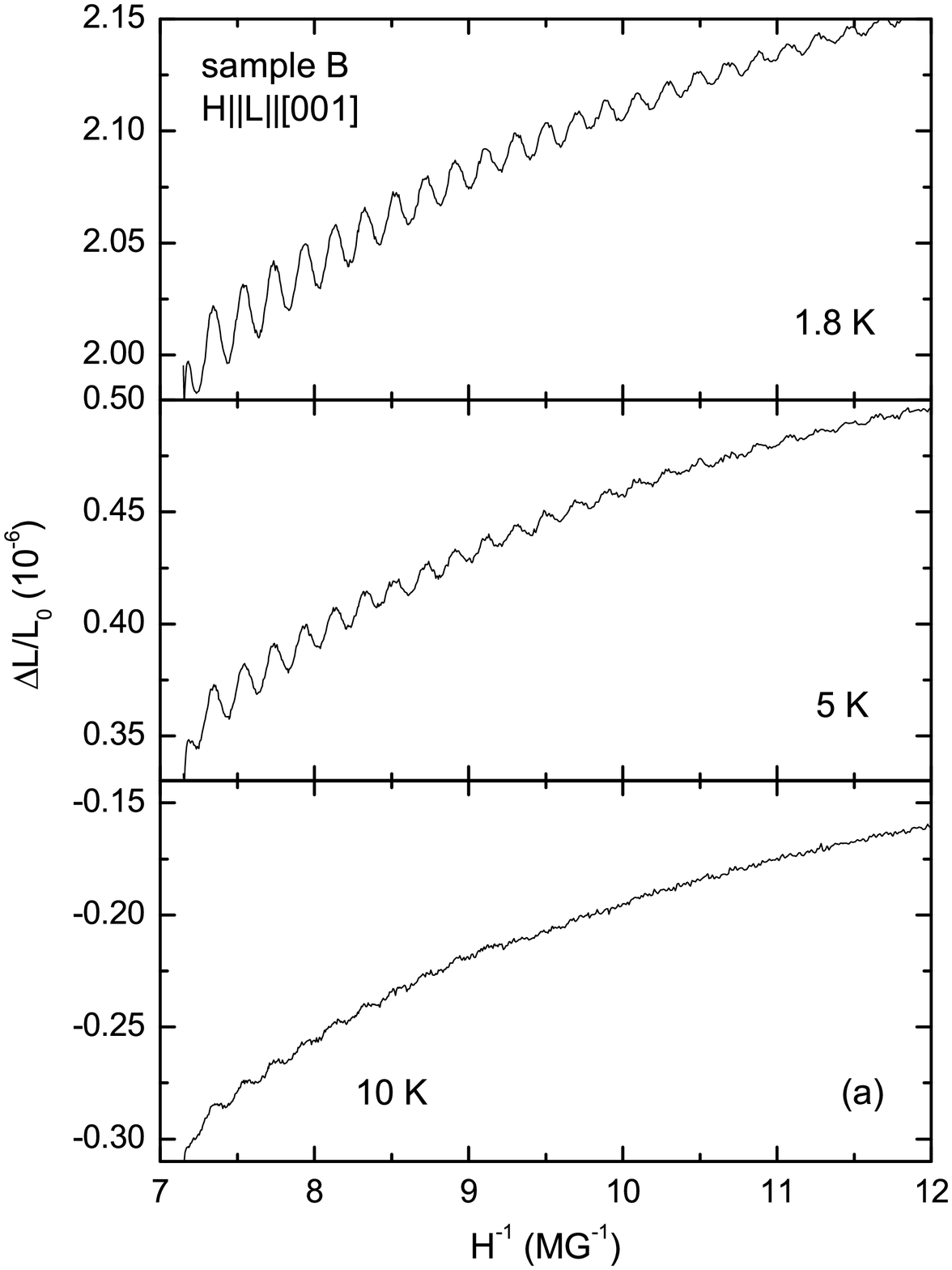}
\includegraphics[angle=0,width=100mm]{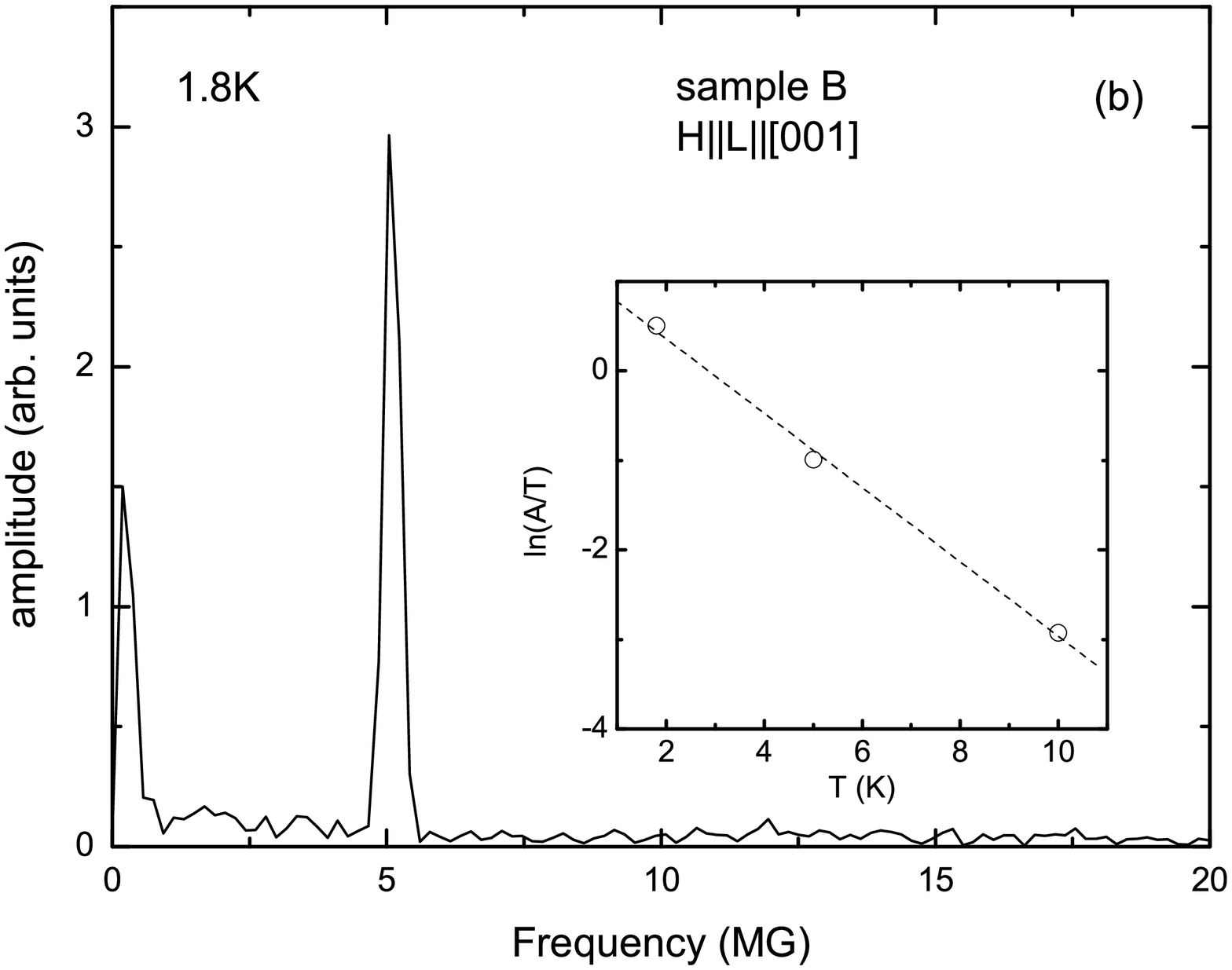}
\end{center}
\caption{(a) Quantum oscillations of the longitudinal magnetostriction plotted as a function of $1/H$ for sample
B. (b) Fourier transform of the $\Delta L/L_0$ vs. $1/H$ curve for $T = 1.8$ K. Inset: $ln(A/T)$ vs. $T$ plot for
amplitude $A$ at $H \sim 133$ kOe.} \label{F8}
\end{figure}

A striking feature in high field magnetostriction of the sample A is clearly seen quantum oscillations (Fig.
\ref{F6}a), that start in the still superconducting, irreversible, region (at approximately 0.7 $H_{c2}$ at $T =
1.8$ K) and continue, with growing amplitude, in higher fields (similarly, the oscillations are present for the
sample B, but the Y-scale in Fig. \ref{F6}b does not allow to see them, see below). Quantum oscillations in
magnetostriction is a known, albeit not so common phenomenon \cite{cha71a,sho84a}, that requires high quality
crystals and sensitive measurement techniques to be observed. These oscillations are present, with smaller
amplitude, at least up to $T = 10$ K and are observed as well in the sample B. Magnetostriction data as a function
of inverse magnetic field for three different temperatures are shown in figures \ref{F7}a (sample A) and \ref{F8}a
(sample B). At base temperature the amplitude of the oscillations for the sample A is $\sim 4$ times higher,
consistent with lower scattering in this sample. The Fourier transformation of the data (figures \ref{F7}b,
\ref{F8}b) reveals one frequency, $F = 5.04$ MG. The effective mass, $m*$, corresponding to this orbit can be
estimated from the slope of the $ln(A/T)$ vs. $T$ plot ($A$ - amplitude of the oscillations in some chosen field)
\cite{sho84a}. For samples A and B this procedure results in values 0.38 $m_0$ and 0.43 $m_0$ respectively; in
average: $m* = 0.4 \pm 0.03 m_0$. Quantum oscillations in magnetization (de Haas - van Alphen effect) in
YNi$_2$B$_2$C were described in number of publications (see \textit{e.g.} \cite{win01a,yam04a,ign05a} and
references therein). The aforementioned oscillations in magnetostriction are consistent, both in frequency and
effective mass values, with the strongest frequency observed by a conventional de Haas - van Alphen effect
measurements, the $\alpha$ orbit on the Fermi surface part formed by the 17th band.

\section{Summary}
In summary, the TE measurements allow us to estimate uniaxial pressure derivatives of $T_c$ which have different
signs for pressure applied in the $ab$ plane and along the $c$ axis. The results call for re-evaluation of the
role of NiB$_4$ structural unit on superconductivity in YNi$_2$B$_2$C. The irreversible MS in YNi$_2$B$_2$C is
large, complex, has features associated with the peak (and/or "dip") effect near $T_c$ and is open for further,
detailed studies of pinning. Quantum oscillations in MS (analog of de Haas - van Alphen effect) were observed for
$H \| c$. The frequency was identified as the $\alpha$ orbit previously seen in magnetization. If combined with
high field magnetization measurements on the same crystals, the oscillatory MS will allow for estimate of
evolution of the extremal orbit area under uniaxial stress \cite{cha71a}.

\ack We thank A. Kreyssig for help in Laue-orienting floating zone grown crystal and M. E. Tillman for writing
software interfacing capacitance bridge with a PPMS environment. We appreciate assistance of B. N. Harmon in
evaluation of Horton Norton and confirmation of our initial results. Ames Laboratory is operated for the U. S.
Department of Energy by Iowa State University under Contract No. W-7405-Eng.-82. This work was supported by the
director for Energy Research, Office of Basic Energy Sciences. One of us (GMS) is supported by the National
Science Foundation under DMR-0305397.

\clearpage

\begin{table}[tbp]
\caption{Changes in specific heat and thermal expansion coefficients at $T_c$ and estimates of anisotropic
pressure derivatives of $T_c$ for two samples of YNi$_2$B$_2$C. Sample A: YNi$_2$$^{10}$B$_2$C, solution grown,
$RRR \approx 63$; sample B: YNi$_2$$^{11}$B$_2$C, melted zone grown, $RRR \approx 16$.} \label{tab}
\begin{tabular}{lccccccc}

\br
Sample&$T_c$&$\Delta C_p$&$\Delta \alpha_{ab}$&$\Delta \alpha_c$&$dT_c/dp_{ab}$&$dT_c/dp_c$&$dT_c/dP*$\\

        &(K)&(mJ/mol K)&($10^{-8}$ K$^{-1}$)&($10^{-8}$ K$^{-1}$)&(K/kbar)&(K/kbar)&(K/kbar)\\
\br
A   &15.9&444&-7.2&5.9&$-1.02 \cdot 10^{-2}$&$0.83 \cdot 10^{-2}$&$-1.21 \cdot 10^{-2}$\\
\mr
B  &15.1&522&-6.5&6.5&$-0.74 \cdot 10^{-2}$&$0.74 \cdot 10^{-2}$&$-0.74 \cdot 10^{-2}$\\
\br

\end{tabular}
\end{table}

\clearpage

\clearpage

\clearpage

\clearpage

\clearpage

\clearpage

\clearpage

\clearpage

\clearpage

\end{document}